\documentclass[aip,12pt]{revtex4}
\usepackage{amsfonts}
\usepackage{graphicx}
\usepackage{dcolumn}
\usepackage{bm}
\usepackage{color}
\usepackage[colorlinks=true,linkcolor=green,anchorcolor=blue,urlcolor=blue,citecolor=blue]{hyperref}
\usepackage{mathrsfs}
\usepackage{graphicx}
\usepackage{mathrsfs}
\usepackage{bbm}
\usepackage{amssymb}
\usepackage{latexsym}
\usepackage{amsmath}
\usepackage{fancyhdr,times}
\usepackage{lastpage}
\usepackage{makeidx}
\usepackage{subfigure}
\usepackage[T1]{fontenc}

\newcommand{\lr}\longrightarrow
\newcommand{\ra}\rightarrow

\newcommand{\gs}{\sigma}
\newcommand{\gth}{\theta}
\newcommand{\gl}{\lambda}
\newcommand{\gd}{\delta}

\begin{document}

\title{\boldmath Leptonic CP Violation and Wolfenstein Parametrization for Lepton Mixing}

\author{Zuo Liu}
\email{liuzhuo@itp.ac.cn}
\author{Yue-Liang Wu}
\email{ylwu@itp.ac.cn}
\affiliation{%
State Key Laboratory of Theoretical Physics (SKLTP)\\
Kavli Institute for Theoretical Physics China (KITPC) \\
Institute of Theoretical Physics, Chinese Academy of Science, Beijing,100190 \\
University of Chinese Academy of Sciences,  P.R.China}

\date{\today}

\begin{abstract}
We investigate a general structure of lepton mixing matrix resulting from
the SU$_F$(3) gauge family model with an appropriate vacuum structure
of SU$_F$(3) symmetry breaking. It is shown that the lepton mixing matrix
can be parametrized by using the Wolfenstein parametrization method to
characterize its deviation from the tri-bimaximal mixing. A general analysis
 for the allowed leptonic CP-violating phase $\delta_e$ and the
 leptonic Wolfenstein parameters  $\lambda_e$, $A_e$, $\rho_e$ is
 carried out based on the observed lepton mixing angles. We demonstrate
 how the leptonic CP violation correlates to the leptonic Wolfenstein
 parameters.  It is found that the phase $\delta_e$ is strongly constrained and
 only a large or nearly maximal leptonic CP-violating
 phase $|\delta_e| \simeq 3\pi/4 \sim \pi/2$ is favorable
  when $\lambda_e > 0.15 $. In particular, when taking $\lambda_e$ to be the Cabbibo angle $\gl_e\simeq \lambda \simeq 0.225$, a sensible result for leptonic Wolfenstein parameters and CP violation is obtained with  $ A_e=1.40$, $\rho_e=0.20$ ,  $\delta_{e}\sim 101.76\;^o$, which is compatible with the one in quark sector. An interesting correlation between leptons and quarks is observed, which indicates a possible common origin of masses and mixing for the charged-leptons and quarks.
\end{abstract}


\maketitle

\newpage

\section{Introduction}

The standard model (SM) has been well established with the observation of the last
 particle predicted in the SM, i.e., Higgs particle, at the LHC experiment\cite{ATLAS,CMS}.
The neutrino oscillations with massive
neutrinos\cite{SuperK,SNO,KamLAND,Soudan,MARCO,K2K,SK,GNO,CHOOZ,T2K,MINOS}
provide a strong evidence and a useful window for exploring new
physics beyond the SM.  In comparison with the quark masses and
CKM quark mixing matrix\cite{CKM}  in the SM, the smallness of
neutrino masses and large MNSP lepton mixing\cite{MNSP} have been
a long-term puzzle to be understood as a possible indication for
new physics. The greatest success of the SM is the gauge symmetry
structure $SU_{c}(3)\times SU_{L}(2)\times U_{Y}(1)$ which
characterizes three basic forces of strong and electroweak
interactions. All the gauge symmetries are associated with the
quantum numbers of quarks and leptons. $SU_c(3)$ characterizes the
symmetry among three color quantum numbers of quarks, $SU_L(2)$
describes the symmetry between two isospin quantum numbers of
quarks and leptons for each family, and $U_Y(1)$  is the symmetry
corresponding to the hypercharge quantum number of quarks and
leptons. The quark and lepton mixing matrices and CP violations
reflect the properties of three family quarks and leptons. To
understand the quark and lepton mixing matrices and CP violations,
it is interesting to investigate the possible gauge symmetries
among three family quantum numbers. Obviously, a non-abelian gauge
family
symmetry\cite{Wu_su3_2012,Wu:2008zzj,Wu:2008ep,YLW1,YLW2,YLW3,YLW4,YLW5,YLW6,CS,MA,CW,BHKR}
for three families of quarks and leptons becomes natural as a
simple extension of the SM gauge symmetry structure.

It has been shown in ref. \cite{Wu_su3_2012}  that the SU$_F$(3) gauge
 family symmetry enables us to construct a simple gauge family model for understanding
 the mixing and masses of leptons. The SU$_F$(3) gauge family symmetry was first
 introduced in early time for estimating the top quark mass\cite{Yanagida:1979gs}.
 It was found in ref. \cite{Wu_su3_2012} that the model can provide a consistent
 prediction for the lepton mixing and neutrino masses when
 considering the appropriate vacuum structure of SU$_F$(3) gauge symmetry breaking. Specifically, through
 appropriately making the SU$_F$(3) gauge fixing condition with keeping a
 residual $Z_2$-permutation symmetry in the neutrino sector, we can obtain
 in the neutrino sector the so-called tri-bimaximal mixing
 matrix\cite{TBM,TBM1,TBM2,TBM3,TBM4} and largely degenerate neutrino masses,
 while the small mixing matrix in the charged-lepton sector is resulted by
 requiring the vacuum structure of spontaneous symmetry breaking to possess
 approximate global U(1) family symmetries\cite{Wu_su3_2012}. Thus the deviation
 from tri-bimaximal mixing in the lepton mixing matrix is attributed to the small
 mixing in the charged-lepton sector, its smallness is protected by the
 mechanism of approximate global U(1) family symmetries
 \cite{HW,Wu:1994ja,Wolfenstein:1994jw,Wu:1994vx}. As the spontaneously symmetry breaking
  CP-violating phases in the vacuum\cite{TDL} are not restricted by the considered
  symmetries, they can in principle be large and maximal. The small masses of the
  neutrinos and charged leptons are simply ascribed to the usual seesaw mechanism.
   As a simple case, when applying the Wolfenstein parametrization\cite{LW} for the CKM quark mixing matrix to the charged-lepton mixing matrix with a similar hierarchy structure as the CKM quark mixing matrix, and making a naive ansatz that all the smallness due to the approximate global U(1) family symmetries is characterized by a single Wolfenstein parameter $\lambda \simeq 0.22$,  we can obtain an interesting prediction for the lepton mixing matrices with a maximal spontaneous CP violation $ \delta \simeq \pi/2$\cite{Wu_su3_2012}
\begin{eqnarray}
& & \sin^2\theta_{13} \simeq \frac{1}{2} \lambda^2 \simeq 0.024\, \quad (\ \mbox{or}\, \ \sin^22\theta_{13} \simeq 0.094), \\
& & \sin^2\theta_{12} \simeq \frac{1}{3},\quad \sin^2\theta_{23} \simeq \frac{1}{2},
\end{eqnarray}
which agrees with the current experimental data\cite{NO_DayaBay_An:2012eh,NO_RENO_Ahn:2012nd, PDG_2012_PhysRevD.86.010001}. The corresponding Leptonic Jarlskog CP-violating invariant quantity\cite{JCP} reaches the maximal value
\begin{eqnarray}
J_{CP}^e \simeq \frac{1}{6}\lambda\sin\delta \simeq 0.037\; .
\end{eqnarray}
The resulting neutrino masses are largely degenerate with the value at the order $m_{\nu_i} \simeq O(\lambda^2) \simeq 0.04\sim 0.06$ eV with a total mass $\sum m_{\nu} \sim 0.15$ eV, which is much larger than the minimal limit $\sum m_{\nu} \sim 0.05$ eV and is expected to be tested by the future experiments.

It is widely expected that the leptonic CP violation can be maximal or large enough to account for the observed matter-antimatter asymmetry in the universe via the leptogenesis mechanism as the CP violation in the SM is not enough to understand the baryogenesis. In this note, we are going to make a general analysis on the leptonic CP-violating phase and its correlation with the deviation from the tri-bimaximal neutrino mixing based on the current experimental results.

\section{Wolfenstein Parametrization of Lepton Mixing Matrix for Characterization of Deviation From Tri-bimaximal Mixing}

Let us begin with the following general structure of MNSP lepton mixing matrix
\begin{eqnarray}
& & V_{MNSP}  = V_e^{\dagger} V_{\nu}   \nonumber \\
         \quad         & & \equiv    P_{\beta} \left(
                     \begin{array}{ccc}
                       c_{12}c_{13}\ \ & s_{12}c_{13}\ \ & s_{13}e^{-i\delta} \\
                       -s_{12}c_{23}-c_{12}s_{23}s_{13}e^{i\delta}\  \ &
                       c_{12}c_{23} - s_{12}s_{23}s_{13}e^{i\delta} \ \ & s_{23} c_{13} \\
                       s_{12} s_{23} - c_{12}c_{23}s_{13}e^{i\delta} \ \
                       & -c_{12}s_{23} - s_{12}c_{23}s_{13} e^{i\delta}\  \
                       & c_{23}c_{13} \\
                     \end{array}
                   \right) \left(
             \begin{array}{ccc}
                 e^{i\phi_1} & 0 & 0 \\
               0 &  e^{i\phi_2} & 0 \\
               0 & 0 & 1 \\
             \end{array}
           \right), \label{eq:St_form}
\end{eqnarray}
which has been expressed into the standard form with  $c_{ij}=cos\theta_{ij},\; s_{ij}=sin\theta_{ij}$\cite{PDG_2012_PhysRevD.86.010001}.  An interesting symmetric parametrization was discussed in\cite{SP}. Where $P_{\beta}$ is a diagonal matrix of phase factors and can be rotated away by the redefinition of charged lepton fields, and $\phi_i$ are the so-called Majorana phases for Majorana neutrinos. It is known that the lepton mixing matrix generally arises from two mixing matrices $V_e$ and $V_{\nu}$, they correspond to the charged-lepton and neutrino mixing matrices arising from diagonalizing the charged-lepton mass matrix and neutrino mass matrix respectively. When the charged-lepton mass matrix is Hermitian $M_e = P_{\delta} U_e m_E U_e^{\dagger} P_{\delta}^{\dagger} = M_e^{\dagger}$ with $m_E$ the diagonal mass matrix of charged leptons $m_E = diag.(m_e, m_{\mu}, m_{\tau} )$, the unitary charged-lepton mixing matrix $V_e$ can in general be written as
 \begin{eqnarray}
 V_e^{\dagger} & = &  U_e^{\dagger}  P_{\delta}^{\ast} \nonumber \\
 & & U_e^{\dagger} \equiv
              \left(
                     \begin{array}{ccc}
                       c_{12}^ec_{13}^e\ \ & s_{12}^ec_{13}^e\ \ & s_{13}^e e^{-i\delta'_e} \\
                       -s_{12}^ec_{23}^e-c_{12}^es_{23}^es_{13}^e e^{-i\delta'_e}\  \ &
                       c_{12}^ec_{23}^e - s_{12}^es_{23}^es_{13}^e e^{-i\delta'_e} \ \ & s_{23}^e c_{13}^e \\
                       s_{12}^e s_{23}^e - c_{12}^ec_{23}^es_{13}^e e^{-i\delta'_e}\ \ & -c_{12}^es_{23}^e - s_{12}^ec_{23}^es_{13}^e e^{-i\delta'_e}\  \
                       & c_{23}^ec_{13}^e \\
                     \end{array}
                   \right), \\
                & &    P_{\delta}  =  \left(
             \begin{array}{ccc}
               e^{i\delta_1} & 0 & 0 \\
               0 &  e^{i\delta_2}  & 0 \\
               0 & 0 &  e^{i\delta_3}  \\
             \end{array}
           \right),
\end{eqnarray}
where $U_e$ is a unitary matrix with CP-violating phase $\delta'_e$, and mixing angles $c_{ij}^e =cos\theta_{ij}^e,\; s_{ij}^e = sin\theta_{ij}^e $. $P_{\delta}$ is a diagonal phase matrix with three phases $\delta_i$, while only two relative phases $(\delta_i-\delta_j)$ are physically observable CP-violating phases.

 For the neutrino mixing matrix, when an appropriate $Z_2$-symmetric neutrino mass matrix between the second and third neutrinos is considered to have three independent matrix elements, the resulting neutrino mixing matrix is completely determined to be
\begin{eqnarray}
 V_{\nu} = \left(
  \begin{array}{ccc}
    \frac{2}{\sqrt{6}}
     & \frac{1}{\sqrt{3}} & 0 \\
    -\frac{1}{\sqrt{6}} & \frac{1}{\sqrt{3}} & \frac{1}{\sqrt{2}} \\
    -\frac{1}{\sqrt{6}} & \frac{1}{\sqrt{3}} & -\frac{1}{\sqrt{2}} \\
  \end{array}
\right),
\end{eqnarray}
which is the so-called tri-bimaximal neutrino mixing matrix.

 It is noticed that when three phases $\delta_{ij}^m$ ($i,j =1,2,3, i< j$) in the Hermitian mass  matrix are not independent and they are related via $\delta_{ij}^m= \delta_i - \delta_j$, one then has
 \begin{equation}
  \delta'_e = 0\; .
\end{equation}
On the other hand, when $\theta_{13}^e$ is small in comparison with $\theta_{12}^e$ and $\theta_{23}^e$, i.e., $\theta_{13}^e \ll \theta_{12}^e, \theta_{23}^e$,  it is easily seen that $\delta'_e$ will not be a dominant source of leptonic CP violation as the CP-violating phase $\delta'_e$ is associated with the mixing angle $\theta_{13}^e$.  In this situation, we may neglect the effect of CP-violating phase $\delta'_e$ and take a typical case $\delta'_e \simeq 0$ for simplicity of discussions. With these considerations, we may replace the unitary matrix $U_e$ by an orthogonal rotation matrix $O_e = U_e (\delta'_e =0)$
 \begin{eqnarray}
 U_e^{\dagger} \to O_e^{T} =
              \left(
                     \begin{array}{ccc}
                       c_{12}^ec_{13}^e\ \ & s_{12}^ec_{13}^e\ \ & s_{13}^e  \\
                       -s_{12}^ec_{23}^e-c_{12}^es_{23}^es_{13}^e \  \ &
                       c_{12}^ec_{23}^e - s_{12}^es_{23}^es_{13}^e \ \ & s_{23}^e c_{13}^e \\
                       s_{12}^e s_{23}^e - c_{12}^ec_{23}^es_{13}^e \ \ & -c_{12}^es_{23}^e - s_{12}^ec_{23}^es_{13}^e \  \
                       & c_{23}^ec_{13}^e \\
                     \end{array}
                   \right).
\end{eqnarray}
To investigate whether the leptonic CP violation can be maximally large with the present experimental measurements on the three mixing angles and the lepton mixing matrix can be characterized by the Wolfenstein parametrization method, we may make a sensible analysis by simply taking $\delta'_e =0$. In ref.\cite{constraint1}, the angle $\theta_{13}^e$ is assumed to be zero, thus the effect of $\delta'_e$ automatically disappears.  An alternative consideration was analyzed in\cite{constraint2}, where the phase $\delta'_e$ was assumed to be the only CP-violating source and the phases $\delta_i$ are taken to be zero, i.e., $\delta_i =0$.

 The leptonic mixing angles $\theta_{ij}$ and mass-square differences $\Delta m^{2}_{ij}$ have been measured by many experiments including the solar neutrino experiment, atmospheric neutrino experiment, accelerator experiment and reactor experiment. The best-fit results presented in PDG\cite{PDG_2012_PhysRevD.86.010001} are
\begin{eqnarray}
\label{eq:para_pdg}
 \sin^{2}2\theta_{12}&=&0.857\pm0.024 ,\notag\\
 \sin^{2} 2\theta_{13}&=&0.095\pm 0.010,\notag\\
 \sin^2 2\theta_{23}&>&0.95.
\end{eqnarray}
which slightly deviates from the tri-bimaximal neutrino mixing.

Note that the presently extracted mixing angles from experiments are not sensitive to the CP-violating phase due to the smallness of the effects concerning the CP violation. As the leptonic CP violation is strongly correlated to the non-zero $\theta_{13}$  which characterizes the deviation from the tri-bimaximal neutrino mixing matrix,  it is then interesting to investigate the leptonic CP-violating phase and its correlation with the deviation from the tri-bimaximal neutrino mixing based on the above structure of lepton mixing matrix and the current experimental results. It is seen that the deviation from tri-bimaximal neutrino mixing is described by the charged-lepton mixing matrix $U_e$ or orthogonal rotation matrix $O_e$, the smallness of the mixing angle $\theta_{13}$ indicates that $s_{12}^e \sim \mathcal{O}(0.1)$, which motivates us to parameterize the rotation matrix $O_e$ via the Wolfenstein parametrization\cite{LW} with a hierarchy structure similar to the CKM quark mixing matrix. With the leptonic Wolfenstein parameter $s_{12}^e \sim \lambda_e\sim \mathcal{O}(0.1)$,  the charged-lepton mixing matrix can be written, to the order $\mathcal{O}(\lambda_{e}^3)$, as the following form:
\begin{eqnarray}
  V_e^{\dagger} \simeq P_\delta^*
   \left(
\begin{array}{ccc}
 1-\frac{\lambda_{e} ^2}{2} &  \lambda_{e}\;  e^{i(\text{$\delta_1$}-\text{$\delta_2$} )} &
 A_{e} \lambda_{e} ^3 \rho_{e}\;  e^{i (\text{$\delta_1$}-\text{$\delta_3$}) } \\
 -\lambda_{e}\;  e^{i (\text{$\delta_2$}-\text{$\delta_1$} ) }   &
   1-\frac{\lambda_{e} ^2}{2} & A_{e}  \lambda_{e} ^2\;  e^{i (\text{$\delta_2$}-\text{$\delta_3$} ) } \\
 A_{e} \lambda_{e} ^3 (1-\rho_{e}) \;  e^{i (\text{$\delta_3$}-\text{$\delta_1$} ) }  & -A_{e} \lambda_{e} ^2\; e^{i (\text{$\delta_3$}-\text{$\delta_2$} ) }  &  1 \\
\end{array}
\right),
  \end{eqnarray}
where the phase matrix $P_\delta^*$ can be absorbed by the redefinitions of charged lepton fields. Note that there is no corresponding Wolfenstein parameter $\eta_e$ in the above parametrization as we have neglected the CP-violating phase $\delta'_e$.
Thus the lepton mixing matrix is given by
\begin{eqnarray}
\label{eq:MSNP_1}
  V_{MSNP} =
   \left(
\begin{array}{ccc}
 1-\frac{\lambda_{e} ^2}{2} &  \lambda_{e}\;  e^{i\text{$\delta_{12}$}} &
 A_{e} \lambda_{e} ^3 \rho_{e}\;  e^{i \text{$\delta_{13}$} } \\
 -\lambda_{e}\;  e^{-i \text{$\delta_{12}$}}   &
   1-\frac{\lambda_{e} ^2}{2} & A_{e}  \lambda_{e} ^2\;  e^{i \text{$\delta_{23}$}} \\
 A_{e} \lambda_{e} ^3 (1-\rho_{e}) \;  e^{-i \text{$\delta_{13}$} }  & -A_{e} \lambda_{e} ^2\; e^{-i \text{$\delta_{23}$}}  &  1 \\
\end{array}
\right)
  \left(
  \begin{array}{ccc}
    \frac{2}{\sqrt{6}}
     & \frac{1}{\sqrt{3}} & 0 \\
    -\frac{1}{\sqrt{6}} & \frac{1}{\sqrt{3}} & \frac{1}{\sqrt{2}} \\
    -\frac{1}{\sqrt{6}} & \frac{1}{\sqrt{3}} & -\frac{1}{\sqrt{2}} \\
  \end{array}
\right),
\end{eqnarray}
which shows that based on the tri-bimaximal neutrino mixing, the lepton mixing matrix can be parametrized by three leptonic Wolfenstein parameters: $\lambda_e,\; A_e,\; \rho_{e},\;$, and CP-violating phases $\delta_{ij}=\delta_i - \delta_j $ ($i=1,2,3$) with $\delta_{23} = \delta_{21}- \delta_{31}$.

As indicated from $Z_2$ symmetry of vacuum structure in the $SU(3)_F$ model\cite{Wu_su3_2012},  it is reasonable to assume that $\delta_2\simeq \delta_3$. When expressing the lepton mixing matrix $V_{MNSP}$ to be the standard form by requiring the matrix elements $V_{11}, V_{12}, V_{23}, V_{33}$ be real with keeping two independent Majorana phases, we can read off the leptonic CP-violating phase from $V_{13}$
\begin{eqnarray}
 \delta_e =\delta_2-\delta_1\simeq \delta_3-\delta_1.
\end{eqnarray}
The Wolfenstein parametrization of lepton mixing matrix is simplified to be
\begin{eqnarray}
\label{eq:MSNP_2}
  V_{MSNP} =
  \left(
\begin{array}{ccc}
 1-\frac{\lambda_{e} ^2}{2} & \lambda_{e} e^{-i\text{$\delta_e$}}   & A_{e}  \lambda_{e} ^3 \rho_{e} e^{-i \text{$\delta_e$}}   \\
 -\lambda_{e}  e^{i \text{$\delta_e$}}  & 1-\frac{\lambda_{e} ^2}{2} & A_{e} \lambda_{e} ^2 \\
 A_{e} \lambda_{e} ^3 (1-\rho_{e}  )e^{i \text{$\delta_e$} } & -A_{e} \lambda_{e} ^2 & 1 \\
\end{array}
\right)
  \left(
  \begin{array}{ccc}
    \frac{2}{\sqrt{6}}
     & \frac{1}{\sqrt{3}} & 0 \\
    -\frac{1}{\sqrt{6}} & \frac{1}{\sqrt{3}} & \frac{1}{\sqrt{2}} \\
    -\frac{1}{\sqrt{6}} & \frac{1}{\sqrt{3}} & -\frac{1}{\sqrt{2}} \\
  \end{array}
\right).
\end{eqnarray}

In terms of the standard form eq.(\ref{eq:St_form}), the lepton mixing matrix can be rewritten in terms of the leptonic Wolfenstein parameters as the following form
\begin{eqnarray}
\label{eq:MSNP_final}
 V_{MSNP}=\left(
\begin{array}{ccc}
  \left|V_{1,1}\right| &
   \left|V_{1,2}\right| & \frac{ \lambda _e}{\sqrt{2}} \left(1-A_e \lambda _e^2 \rho
   _e\right)e^{-i \delta_e} \\
 e^{-i \phi _1} V_{2,1} & e^{-i \phi _2} V_{2,2} &
 \frac{1}{\sqrt{2}} \left(1- A_e \lambda _e^2-\lambda _e^2/2\right)  \\
 e^{-i \phi _1} V_{3,1} & e^{-i \phi _2} V_{3,2} & -\frac{1}{\sqrt{2}} \left(1+ A_e \lambda
   _e^2\right) \\
\end{array}
\right) \left(
\begin{array}{ccc}
 e^{i \phi _1} & 0 & 0 \\
 0 & e^{i \phi _2} & 0 \\
 0 & 0 & 1 \\
\end{array}
\right),
\end{eqnarray}
where $V_{i,j}$ are the matrix elements of  $V_{MNSP}$ via eq.(\ref{eq:MSNP_2}).
The two Majorana phases $\phi_1,\phi_2$ turn out to be
\begin{eqnarray}
 \phi_1&=&\arg V_{1,1}=\arctan\frac{\left(\lambda _e+A_e \lambda_e^3 \rho_e\right) \sin
   \delta _e}{2-\lambda _e^2-\left(\lambda _e+A_e \lambda_e^3 \rho _e\right) \cos \delta_e},\notag\\
 \phi_2&=&\arg V_{1,2}=\arctan \frac{-\left(\lambda _e+A_e \lambda_e^3 \rho _e\right) \sin
   \delta _e}{1-\lambda _e^2/2+\left(\lambda _e+A_e \lambda_e^3 \rho _e\right) \cos \delta_e}.
\end{eqnarray}
and the mixing angle $\theta_{ij}$ can be expressed in terms of the leptonic Wolfenstein parameters as
\begin{eqnarray}
  \label{eq:relation_s13}
s_{13}&=&\frac{\lambda_{e} }{\sqrt{2}}\big| 1-A_{e}\rho_{e}\lambda_{e}^2\big|,\\
\label{eq:relation_s23}%
 s_{23}&=&\frac{1}{\sqrt{2}}\frac{1}{\sqrt{1-s_{13}^2}} \big|1- \frac{\lambda_e^2}{2} - A_e \lambda_{e}^2 \big|, \\
 \label{eq:relation_s12}%
 s_{12}&=& \frac{1}{\sqrt{3}}\frac{1}{\sqrt{1-s_{13}^2}}\big|
1- \frac{\lambda_e^2}{2} + \lambda_e (1 + A_e\rho_e\lambda_e^2) e^{-i\delta_{e}} \big|.
\end{eqnarray}
which shows that the leptonic Wolfenstein parameters $\lambda_e,\; A_e,\; \rho_{e},\;$ and the CP-violating phase $\delta_e$ characterize the lepton mixing with deviation from the tri-bimaximal mixing.

As an illustration, it is interesting to observe that by taking the leptonic Wolfenstein parameters to be the following typical values with a maximal CP-violating phase
\begin{eqnarray}
\label{eq:parameter}
  \lambda_{e}\sim 0.22,\quad A_{e}\sim 1,\quad \rho_{e}\sim 1,\quad
  \delta_{e}=\delta_2-\delta_1=\delta_3-\delta_1\sim \frac{\pi}{2},
\end{eqnarray}
we obtain the predictions for the lepton mixing angles
\begin{eqnarray}
 \sin^{2}2\theta_{12}&\sim&0.901 ,\notag\\
\sin^{2}2\theta_{13}&\sim&0.086 ,\notag\\
\sin^{2}2\theta_{23}&\sim&0.986 ,
\end{eqnarray}
which are consistent with the PDG's best-fit results given in eq. (\ref{eq:para_pdg}) \cite{PDG_2012_PhysRevD.86.010001} at $1\sigma$ level, except a small mismatch of $\theta_{12}$. Such a consistency shows that the leptonic Wolfenstein parameters chosen in eq.(\ref{eq:parameter}) are in the reasonable region of  parameter space.

Alternatively, we may use the PDG's value of $\theta_{12}$ given in eq.(\ref{eq:para_pdg})
$\sin^{2}2\theta_{12}=0.857\pm0.024$  to extract the leptonic CP-violating phase $\delta_{e}$. With other parameters
chosen as eq.(\ref{eq:parameter}), it is easily found that
\begin{eqnarray}
\delta_{e}=(101.94_{+5.90}^{-6.28}\; )^o ,
\end{eqnarray}
which is very close to the maximal CP-violating phase $\delta_e
\sim 0.57 \pi$. The corresponding two Majorana phases with the
input parameters as eq.(\ref{eq:parameter}) are yielded to be
\begin{eqnarray}
  \phi_1\sim 6.7^{o},\quad
  \phi_2\sim -13.3^{o}.
\end{eqnarray}

We shall make a general constraint on the leptonic Wolfenstein parameters $\lambda_e,\; A_e,\; \rho_{e},\;$ and the CP-violating phase $\delta_e$ by a detailed analysis below.

\section{Constraints on Leptonic Wolfenstein Parameters}

As it is shown in previous section that the deviation from tri-bimaximal lepton mixing matrix can be described by three leptonic Wolfenstein parameters $\lambda_{e}, \; A_e,\; \rho_e$. It is seen from eq.(\ref{eq:relation_s13}) and eq.(\ref{eq:relation_s23})  that $\sin\theta_{13}$ depends on  $\lambda_e$ and $A_e\rho_e$, while $\sin\theta_{23}$ relies on $\lambda_e$ and $A_e$. In this section, we shall take the mixing angles $\theta_{13}$ and $\theta_{23}$  indicated from the measurements as the input to provide a general constraint on leptonic Wolfenstein parameters  $\lambda_{e}, \; A_e,\; \rho_e$.

\subsection{Constraints From $\theta_{13}$ }

The precise measurements on $\theta_{13}$ have been carried out by DayaBay Collaboration group\cite{NO_DayaBay_An:2012eh} and RENO Collaboration group\cite{NO_RENO_Ahn:2012nd}. These two experiments measured the
disappearance of $\overline{\nu}_e$ from the reactor.  The $\Delta m^2_{31}$ dominated amplitude is given by
\begin{eqnarray}
  P(\overline{\nu}_e\rightarrow \overline{\nu}_e)=1-\sin^2
  2\theta_{13}\sin^2\frac{\Delta m^2_{31} L}{4E}.
\end{eqnarray}
which shows that the measured results do not sensitively correlate to the values of other two mixing angles $\theta_{12},\theta_{23}$ and CP-violating phase $\delta_e$. From eq.(\ref{eq:relation_s13}), it is seen that $\theta_{13}$ is also insensitive to the CP-violating phase $\delta_e$. Thus we may use the experimental data on $\theta_{13}$ to make constraints on the leptonic Wolfenstein parameters.

Before doing that, it is noticed that when keeping the expansion of lepton mixing matrix to the order $\mathcal{O}(\lambda_e)$, we arrive at the following simple relation
\begin{eqnarray}
  \lambda_{e}\sim \frac{s_{13}}{s_{23}} \simeq 0.23,
\end{eqnarray}
where we have used the best fit values $\sin^2\theta_{13}\sim 0.0225$ and
$\sin^2\theta_{23}\sim 0.42$\cite{PDG_2012_PhysRevD.86.010001} to yield the numerical value $\lambda_{e}\sim0.23$, which is very close to the Wolfenstein parameter of Cabibbo angle $ \sin\theta_{c}= \gl \simeq 0.225$ in quark sector\cite{PDG_2012_PhysRevD.86.010001}. This observation checks the consistence of the assumption that $\lambda_{e}\sim \mathcal{O}(10^{-1})$.

Let us now turn to make a general analysis by adopting the precisely measured mixing angle $\theta_{13}$\cite{PDG_2012_PhysRevD.86.010001}
\begin{eqnarray}
\label{eq:daya_reno_13}
  \sin^{2} 2\theta_{13}=0.096\pm 0.013(\pm0.040)\; \text{at}
  \;1\sigma\;(3\sigma),
\end{eqnarray}
which enables us to constrain the allowed region of the combined leptonic Wolfenstein parameters $A_e\rho_e$ for a given $\lambda_e$.

\begin{figure}[t]
 \centering
\includegraphics[width=2.5in]{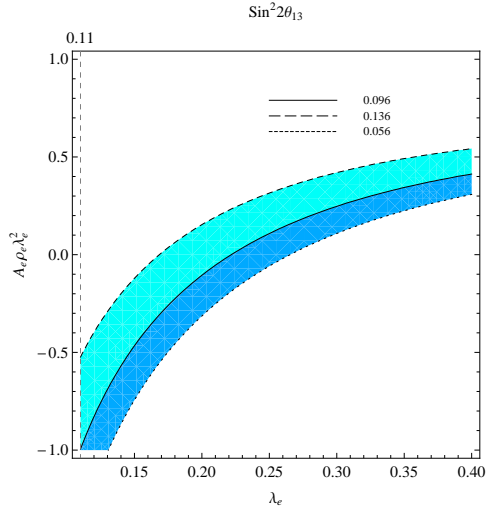}
\caption{The contour plot of $\sin^2 2\theta_{13}$ for
$A_e\rho_e\lambda_e^2$ as a function of $\lambda_e$. The contours
show the best fit value and $3\sigma$ deviations.
}\label{fig:cont_13}
\end{figure}

The contour plot for the input $\sin^{2}2\theta_{13}$ is shown in
figure\ref{fig:cont_13}. It is seen from figure \ref{fig:cont_13}
that $A_e\rho_e\lambda_{e}^{2}<0$ only occur for small values of
$\lambda_e$. In the plot, we have restricted the region to be in the range $-1\leq A_e\rho_e\lambda_e^2 \leq 0.5$, so that it
satisfies the perturbative expanding of Wolfenstein parametrization. It also leads to a reasonable region for the
parameter $\lambda_e$
\begin{equation}
\lambda_e\simeq 0.11 \sim 0.40.
\end{equation}
which will be taken to be a possible allowed region when considering constraints from other two mixing angles $\theta_{12}$ and $\theta_{23}$.

From eq.(\ref{eq:relation_s13}), it is seen that $\sin^2\gth_{13}$
is an even function of $\gl_e$. Thus for the region with
$\gl_e<0$, the contours of $\sin^2\gth_{13}$ are just the mirror
images of figure \ref{fig:cont_13}, which is omitted here.

\subsection{Constraints From $\theta_{23}$  and $\theta_{13}$ }

\begin{figure}[t]
 \centering
\includegraphics[width=2.5in]{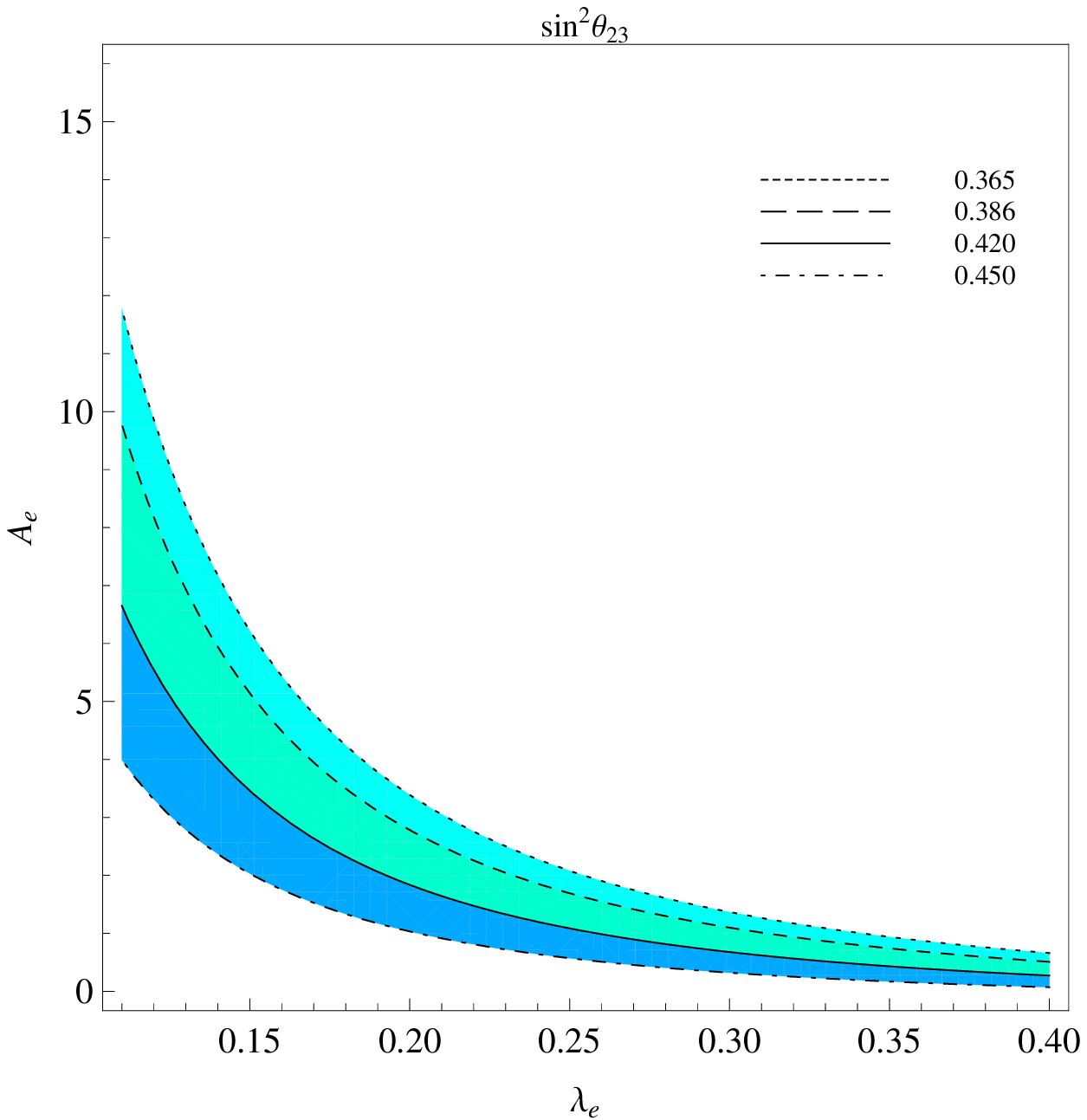}
\includegraphics[width=2.55in]{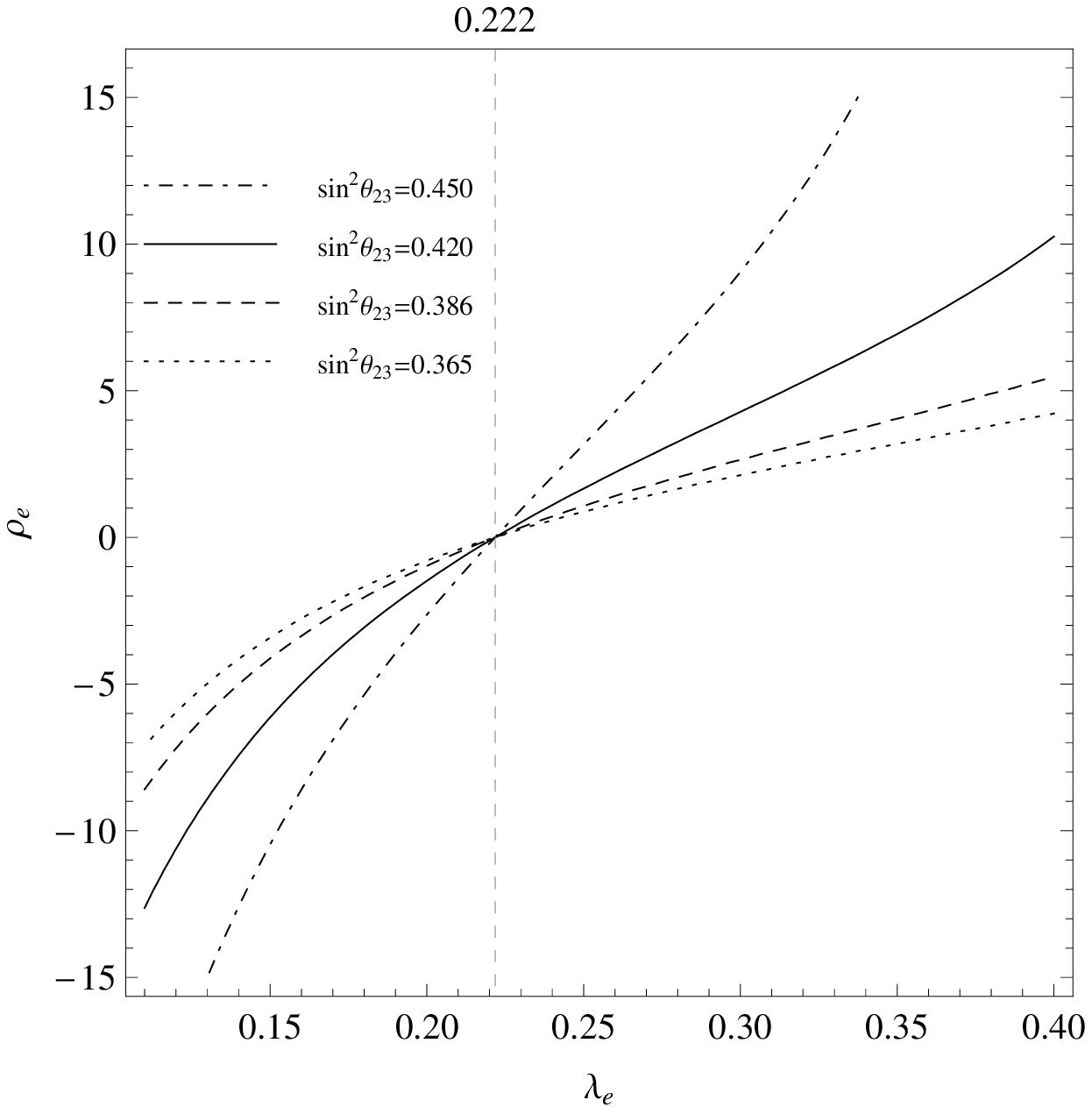}
\caption{The left panel is the contour plots of $\sin^2\theta_{23}$ for $A_e$ as a
function of $\lambda_e$. The right panel is the contour plots of $\sin^2\theta_{23}$ for $\rho_e$ as a function of $\lambda_e$, where  the central value of $\sin^2\theta_{13}$ has been used to fix $A_e\rho_e$ for a given $\lambda_e$. The vertical line labels a critical value of $\lambda_e=\sqrt{2}\sin\theta_{13}$, where $\rho_e=0$.}\label{fig:cont_23}
\end{figure}

For tri-bimaximal mixing, there is a maximal mixing $\sin^2\gth_{23}=1/2$.
The small deviation to the maximal mixing indicates that the leptonic Wolfenstein
parameters should be small, which may still be consistent with the current data within
 the experimental errors. While the recent global fitting results appear to indicate a
 quite large deviation from the maximal mixing with
 $\sin^2\theta_{23}=0.386^{+0.024}_{-0.021}$
 \cite{theta23_1_Fogli:2012ua}, and
 $\sin^2\theta_{23}=0.41^{+0.037}_{-0.025}$
\cite{theta23_2_GonzalezGarcia:2012sz}. For a general discussion,
we may consider a constraint from a wide range of $\theta_{23}$ by
covering over different global fitting results, i.e.,
$\sin^2\gth_{23}\simeq 0.365\sim 0.450$. The resulting constraint
is shown in the left panel of figure \ref{fig:cont_23} for
parameter $A_e$ as a function of $\lambda_e$.

By combining the constraint from $\theta_{13}$ and $\theta_{23}$,
we are able to obtain the constraint for the allowed region of
$\rho_e$ as a function of $\lambda_e$.  As shown in the right
panel of figure\ref{fig:cont_23}, by taking the central value of
$\sin^2\theta_{13}$ given in eq.(\ref{eq:daya_reno_13}), we can
obtain the allowed region for $\rho_e$ as a function of $\gl_e$
from the given values of $\sin^2\theta_{23}$. It is seen that a
wide region swept by the curve when $\sin^2\theta_{23}$ increasing
from $0.365$ to $0.450$ is allowed.

Note that there is a special situation that for $\gl_e=\sqrt{2}
\sin\theta_{13}$, then $A_e\rho_e=0$, namely $\rho_e =0$ for
$A_e\neq 0$. As a consequence, four curves intersect with each
other at this point, as indicated in figure\ref{fig:cont_23}.

\section{Leptonic CP Violation and Lepton-Quark Correlation}

In this section, we should make a general analysis on the leptonic CP-violating phase and its correlation to the deviation from the tri-bimaximal neutrino mixing, which is characterized by the leptonic Wolfenstein parameters as discussed in the previous section.

\subsection{Constraints From $\theta_{12}$ and Leptonic CP Violation}

It is seen from eq. (\ref{eq:relation_s12}) that $\sin\theta_{12}$ depends on CP-violating phase $\delta_e$, $\lambda_e$ ,$A_e\rho_e$.  Here $A_e\rho_e$ can be constrained from $\theta_{13}$ for a given $\lambda_e$.

\begin{figure}[t]
 \centering
\includegraphics[width=2.5in]{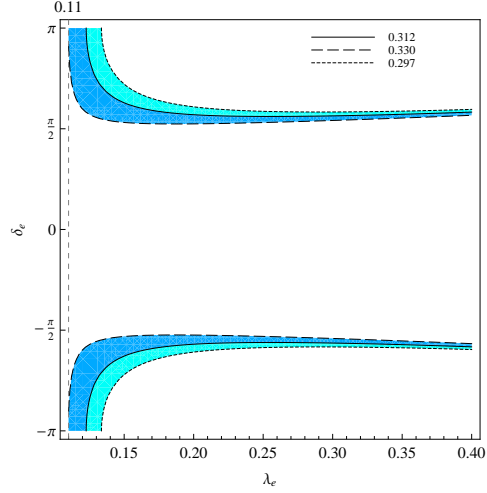}
\caption{The contour plot of $\sin^2\theta_{12}$ for the
CP-violating phase $\delta_e$ as a function of $\lambda_e$ with
the best fit value of $\sin^2\gth_{12}$ and $1\gs$ deviations. For
$0.11\leq\lambda_e\leq 0.15$, the allowed $\delta_{e}$ ranges from $\pm \pi$ to
$\pm \pi/2$, and for $\lambda>0.2$, the resulting $\delta_{e}$ is close to maximal $\pm \pi/2$.}\label{fig:cont_12}
\end{figure}

The mixing angle $\theta_{12}$ is well determined from solar
neutrino oscillation experiments. The measured value of
$\theta_{12}$ generally correlates to the value of $\theta_{23}$.
It is convenient to obtain the values of $\theta_{12}$  by setting
$\sin^2 2\theta_{23}=1$. The global fitting results have provided
us with both values of $\theta_{23}$ and $\theta_{12}$. Although a
non-maximal $\theta_{23}$ is hinted\cite{theta23_1_Fogli:2012ua},
there is no tension among different global fitting results on
$\theta_{12}$. For instance,
$\sin^2\gth_{12}=0.307^{+0.018}_{-0.016}$\cite{theta23_1_Fogli:2012ua}, $\sin^2\gth_{12}=0.311\pm0.013$\cite{theta23_2_GonzalezGarcia:2012sz}, and
$\sin^2\gth_{12}=0.320^{+0.016}_{-0.017}$\cite{global}.
Here we take the result $\sin^2\theta_{12}=0.312^{+0.018}_{-0.015}
$ given in PDG\cite{PDG_2012_PhysRevD.86.010001} to make
constraints on the CP-violating phase $\delta_e$ as a function of
$\lambda_e$. The value of  $A_e\rho_e$ is constrained from
$\theta_{13}$. The allowed region for CP-violating phase
$\delta_e$ is given as a function of $\lambda_e$ in
figure\ref{fig:cont_12}, where we have taken the central value
$\sin^2 2\theta_{13}=0.096$ to yield the value of $A_e\rho_e$.

It is seen from figure \ref{fig:cont_12} that there are two
special regions: for $0.11\leq\lambda_e\leq 0.15$, the values of
$\sin^{2}\gth_{12}$ is insensitive to $\gd_{e}$, the allowed
region of $\delta_e$ ranges from $\pm \pi $ to $\pm \pi/2$. While
for $\lambda_e\geq 0.2$,  the constraint on $\gd_{e}$ becomes very
strong, the resulting CP-violating phase is near maximal
$\gd_{e}\sim\pm \pi/2$. In this region, $\sin^2\theta_{12}$ is
insensitive to the values of $\gl_e$. Thus the leptonic CP
violation favors a maximal CP violation for a large range of
leptonic Wolfenstein parameter $\lambda_e$. Note that a minimal
CP-violating phase $\gd_{e}\sim 1.08\pi$ was obtained in a global
fit \cite{theta23_1_Fogli:2012ua}  when the atmospheric neutrino
data are included, while such a fitting result corresponds to a
special region in the parameter space, which does not exclude a
large or nearly maximal CP violation.

\subsection{Combination of All Constraints and Lepton-quark Correlation}

\begin{figure}[t]
 \centering
\scalebox{0.7}{\includegraphics{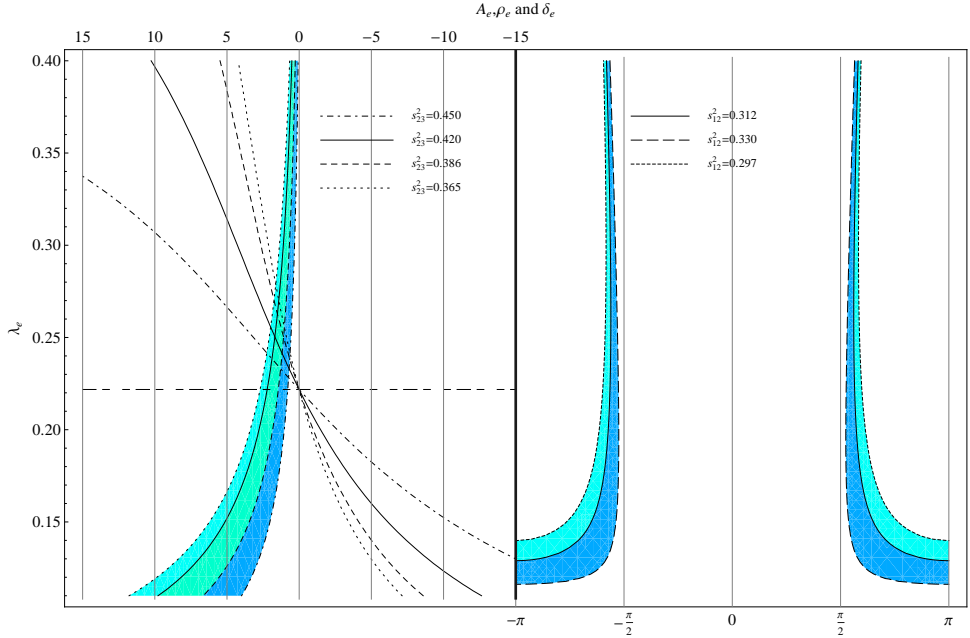}} \caption{The allowed parameter regions for $A_e$,$\rho_e$ and $\gd_{e}$ for given values of $\lambda_e$.} \label{fig:final}
\end{figure}

It is useful to combine all the constraints obtained from $\sin^2\theta_{13}$,  $\sin^2\theta_{23}$ and  $\sin^2\theta_{12}$ and
 plot them together in the same figure\ref{fig:final}, so that it is easily seen the allowed values of $A_e$, $\rho_e$ and $\delta_e$ for a given $\gl_e$.  A big uncertainty arises from whether $\sin^2\theta_{23}$ is largely deviate from the maximal mixing, which makes the allowed values of $A_e$ and $\rho_e$ become large.

It is easy to see from figure\ref{fig:final} that there are two
typical regions for leptonic Wolfenstein parameters characterized
with a small $\lambda_e$ ( $\lambda_e < 0.15$) and a large
$\lambda_e$ ($\lambda_e >0.15 $). For the small $\lambda_e $, we
have
  \begin{eqnarray}
   \gl_e\in [0.11,0.15],\; A_e\in[12, 2],\;  \rho_e\in[-12,-2],\;\;  |\gd_{e}|\in[\pi, \pi/2).
  \end{eqnarray}
which shows that the CP-violating phase $\gd_{e}$ is not well constrained in this case.

A  global fitting result cited in\cite{theta23_1_Fogli:2012ua} corresponds to a solution of the small $\gl_e$ with
$\delta_{e}\sim\pi$. From the results given in\cite{theta23_1_Fogli:2012ua} for the normal hierarchy:
$\sin^2\theta_{23}=0.386^{+0.024}_{-0.021}$,
$\sin^2\theta_{13}=0.0241\pm 0.0025$,
$\sin^2\theta_{12}=0.307^{+0.018}_{-0.016}$, and $\gd_{e}=1.08\pi$,
one can easily read from Fig.\ref{fig:final} the corresponding leptonic Wolfenstein parameters
\begin{eqnarray}
\label{eq:Fogli_NH}
 \gl_{e}=0.127^{-0.013}_{+0.012},\quad
 A_{e}=7.27^{-1.67}_{+1.50},\quad
 \rho_{e}=-6.21^{-0.75}_{+0.79}.
\end{eqnarray}
With the central values, two Majorana phases are found to be very small
\begin{eqnarray}
  \phi_1=-0.29\;^o,\quad
  \phi_2=0.61\;^o.
\end{eqnarray}
For the case of inverted hierarchy, the result is very close to the above one, we shall omit it here.

For large values of $\lambda_e$, we have
  \begin{eqnarray}
   \gl_e\in [0.15, 0.4],\;  A_e\in[7, 0],\;\; \rho_e\in[-10,15], \; \;  |\gd_{e}|\sim [3\pi/4, \pi/2).
  \end{eqnarray}
where the CP-violating phase $\gd_{e}$ is strongly constrained, only a large or nearly maximal  CP violation is favorable.

It is interesting to observe from Fig. \ref{fig:final} that when
taking the value of the leptonic Wolfenstein parameter $\lambda_e$
to be the same as the one in the quark sector, $\gl_e\simeq
\lambda \simeq 0.225$, and fixing the lepton mixing angles to be
the central values $\sin^{2} 2\theta_{12}=0.857$ and
$\sin^{2}\theta_{23}=0.42$, we arrive at a sensible result for the
leptonic Wolfenstein parameters
\begin{eqnarray}
&&\gl_e \simeq 0.225, \quad A_e=1.40 \notag\\
  && \rho_e=0.20,  \quad  \delta_{e}\sim 101.76\;^o  \notag\\
&&\phi_1=6.40^o,\quad \phi_2=-13.56^o.
\end{eqnarray}
which is compatible with the Wolfenstein parameters in quark sector
\begin{eqnarray}
&&\lambda \simeq 0.225, \quad A=0.811 ,\notag\\
  &&\rho_e=0.131, \quad \eta = 0.345 \quad \mbox{or}\quad \delta \simeq
  69^o .
\end{eqnarray}
In this case, the resulting lepton mixing matrix is given by
\begin{eqnarray}
 V_{MSNP} =\left(
\begin{array}{ccc}
 0.820 & 0.551 &0.157 e^{i0.57\pi} \\
 -0.407-0.135 i & 0.642+0.024 i & 0.639 \\
 -0.378+0.052 i & 0.518+0.132 i & -0.757 \\
\end{array}
\right) \left(
\begin{array}{ccc}
 e^{0.11 i} & 0 & 0 \\
 0 & e^{-0.24i} & 0 \\
 0 & 0 & 1 \\
\end{array}
\right).
\end{eqnarray}

\begin{figure}[t]
 \centering
\scalebox{1.0}{\includegraphics{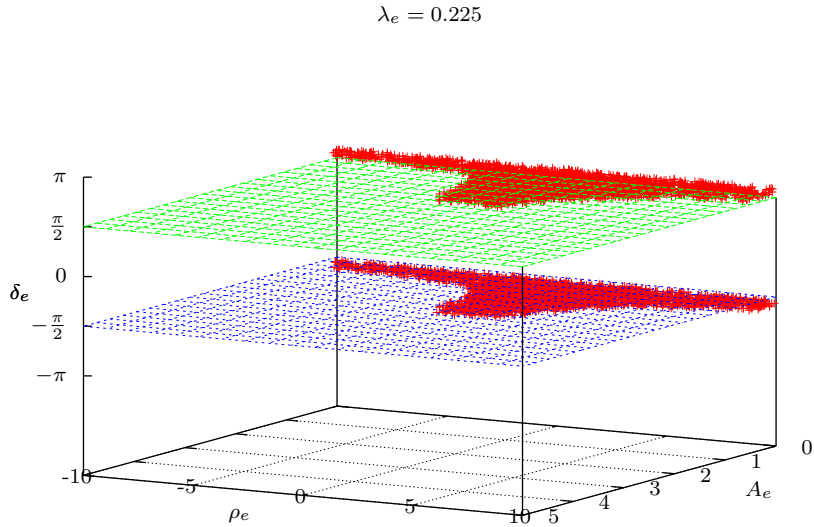}} \caption{The allowed
regions of leptonic Wolfenstein parameters in parameter space for$\lambda_e=0.225$.}
\label{fig:para_space}
\end{figure}

To show manifestly such an interesting situation, it is useful to
plot the leptonic Wolfenstein parameters in a parameter space as
shown in Fig.\ref{fig:para_space} . It is easily seen that when fixing the parameter
$\lambda_e \simeq \lambda \simeq 0.225$, the whole parameters
space for $A_e,\;\rho_e$ and $\delta_{e}$ is almost located on two
planes with $\delta_e\sim \pm \pi/2$. The above analysis implies
that only a large or nearly maximal leptonic CP violation is favorable
in a large region of parameter space when $\lambda_e > 0.15$.

The above results indicate a strong correlation between charged-leptons and quarks. An assumption that $V_e \simeq V_{CKM}$ and $ V_{MNSP} \simeq V_{CKM}^{\dagger} V_{TB}$ was discussed early in\cite{PR,AD,KN}.

\section{Conclusions and Remarks}
\label{sec:conclusion}

We have shown that the lepton mixing can be parametrized by the Wolfenstein parametrization method based on a general structure of lepton mixing matrix, where the mixing matrix from neutrino sector is a tri-bimaximal mixing and the mixing matrix from charged-lepton has small mixing. Such a structure of lepton mixing has been shown to be resulted from the SU$_F$(3) gauge family model\cite{Wu_su3_2012} when considering the appropriate vacuum structure of SU$_F$(3) gauge symmetry breaking. Where the tri-bimaximal mixing can be yielded from the residual $Z_2$-permutation symmetry in the neutrino sector and the small mixing in the charged-lepton sector is led by requiring the vacuum structure of spontaneous symmetry breaking to possess approximate global U(1) family symmetries. We have demonstrated that the small mixing matrix in the charged-lepton sector characterizes  the deviation from tri-bimaximal mixing in the lepton mixing matrix, and can be parametrized by the Wolfenstein parametrization method. As the spontaneous CP-violating phases in the vacuum are in general not restricted by the considered symmetries, so that they can in principle be large and maximal.

Based on the input values of lepton mixing angles $\theta_{13}$, $\theta_{23}$ and $\theta_{12}$ indicated from various neutrino experiments, we have made a general analysis for the allowed leptonic CP-violating phase $\delta_e$ and leptonic Wolfenstein parameters $\lambda_e$, $A_e$, $\rho_e$.  It has explicitly been shown how the leptonic CP violation correlates to the leptonic Wolfenstein parameters which characterize the deviation of tri-bimaximal lepton mixing. For a reasonable range of parameter $\lambda_e \simeq 0.11\sim 0.40$, there appear two typical regions, i.e., one with $\lambda_e \simeq 0.11\sim 0.15$, and other with $\lambda_e \simeq 0.15\sim 0.40$. For the small values of $\lambda_e \simeq 0.11\sim 0.15$, the mixing angles $\gth_{ij}$ are insensitive to $\gd_{e}$, thus the CP-violating phase $\delta_e$ is not well constrained, its allowed region can range from $|\delta_e| \sim \pi$ to $|\delta_e| \sim \pi/2$. While for the large values of $\lambda_e \simeq 0.15\sim 0.40$, the CP-violating phase $\delta_e$ has strongly been constrained, only a large or nearly maximal leptonic CP violation with $|\delta_e| \simeq 3\pi/4\sim \pi/2$ is allowed.

It has been demonstrated that when taking the leptonic Wolfenstein parameter $\lambda_e$ to be the Cabbibo angle in quark sector, $\gl_e\simeq \lambda \simeq 0.225$, we are able to obtain a sensible result with $\gl_e \simeq 0.225$, $ A_e=1.40$, $\rho_e=0.20$ ,  $\delta_{e}\sim 101.76\;^o$, which is compatible with the Wolfenstein parameters in quark sector: $\lambda \simeq 0.225$, $A=0.811$, $\rho_e=0.131$, $ \delta \simeq 69^o$. Such a correlation implies a possible common origin of masses and mixing angles for the charged-leptons and quarks.

In conclusion, the lepton mixing matrix can well be characterized by leptonic Wolfenstein parameters in the basis of tri-bimaximal neutrino mixing. The leptonic CP violation has a strong correlation to the leptonic Wolfenstein parameters, a large or nearly maximal leptonic CP violation is favorable in a large region of parameters. More precise measurements for the lepton mixing angles are very helpful. It is essential to have a direct measurement for the leptonic CP violation in near future.

\acknowledgments

The authors would like to thank Petcov and Valle for useful discussions and comments. This work is supported in part by the National Nature Science Foundation of China (NSFC) under Grants No. 10975170, No.10905084,No.10821504; and the Project of Knowledge Innovation Program (PKIP) of the Chinese Academy of Science.





\end{document}